\documentclass[12pt]{iopart}
\begin{document}

\title[Low energy quantum gravity effects]{On low energy quantum gravity induced
effects on the propagation of  light.}

\author{Reinaldo J Gleiser, Carlos N Kozameh and Florencia Parisi}

\address{Facultad de Matem\'atica, Astronom\'\i a y F\'\i sica, \\ Universidad
Nacional de C\'ordoba, \\ Ciudad Universitaria, (5000) C\'ordoba,
Argentina}

\begin{abstract}

Present models describing the interaction of quantum Maxwell and gravitational fields predict a breakdown of Lorentz invariance and a non standard dispersion relation in the semiclassical approximation. Comparison with observational data however, does not support their predictions. In this work we introduce a different set of ab initio assumptions in the canonical approach, namely that the homogeneous Maxwell equations are valid in the semiclassical approximation, and find that the resulting field equations are Lorentz invariant in the semiclassical limit. 

 We also include a phenomenological analysis of possible
effects on the propagation of light, and their dependence on energy, in a
cosmological context.

\end{abstract}

\submitto{\CQG}
\ead{gleiser@fis.uncor.edu}
\maketitle

\section{Introduction}

Present models describing the interaction of quantum Maxwell and gravitational fields predict a breakdown of Lorentz invariance. In the semiclassical approximation, the common feature in these models is a non standard dispersion relation which shows that the spacetime behaves as a media with a frequency dependent index of refraction \cite{A-Camelia, GaPu, Alfaro, Thiemann}. Another consequence is the selection of a preferred reference frame, namely, the one at rest with respect to the media. In geometrical terms this can be realized by the introduction of a preferred timelike vector field $t^a$ that serves as a universal time.

It is instructive to analyze in pure phenomenologial terms the effect of a semiclassical gravity state on the propagation of electromagnetic radiation. The effective interaction is realized by adding to the standard Maxwell equations extra terms with minimal coupling between the Maxwell field and the above mentioned timelike vector field. 
Although the interaction contains parity breaking and parity preserving terms,
it is perhaps surprising that the only possible mesurable effect arising from our approach comes from the parity breaking interaction and it is this effect, precisely, what is predicted by the leading models on the canonical approach\cite{GaPu, Alfaro, Thiemann}. The main prediction of these models is a dispersion relation that depends on the helicity and the energy of the radiation field. Our analysis also seems to rule out wave propagation that is parity invariant, and in particular recent results by John Ellis, et. al \cite{Ellis}.

Note that if the dispersion relation depends on the helicity then a linearly polarized  wave packet with a continuum spectra will rotate its polarization direction as it travels through space. Since the rotation depends on the energy, the wave will become totally depolarized after traveling a suitable optical path. For light coming from a cosmological source, the observation of linear polarization can be used to set a severe bound on the phenomenological coupling constant of the Gambini-Pullin model \cite{GleiKo} and, as we will show in this work, essentially rule out the Sahlmann-Thiemann construction.

At this point one is tempted to ask whether or not Lorentz invariance is necessarily broken by quantum gravity. Since observational evidence seems to support Lorentz invariance, we reexamine the canonical models looking for assumptions introduced in their construction that, although natural at first sight, might not be true in the final theory. Our hope is that by suitably changing them we may preserve Lorentz invariance.

In particular, in what follows we replace their ab-initio assumption of the electric field and potential being conjugate variables by a different anzatz, namely, that the quantum Maxwell field is the curvature of the Maxwell connection. This natural assumption has the expected consequence that the resulting field equations are Lorentz invariant. Although the calculation is done at a linear level, it is very easy to extend the results to any order in the perturbation parameter. We thus conjecture that Lorentz invariance will still hold in the quantum interaction between gravitons and photons. 

The work is organized as follows. In Section 2 we present a phenomenological approach to the propagation equations showing that if Lorentz invariance is broken, at most there will be a gravity induced rotation of the polarization vector. We also show that the available observational data essentialy rules out the Sahlmann-Thiemann model.

In Section 3 we assume that the Maxwell field is the curvature of a connection, namely, $F = dA$. Since this statement is independent of the existence of a quantum gravitational field we conjecture that it is valid for the full theory. Using the energy density of the electromagnetic field and this new assumption, we obtain different conjugate variables and thus different field equations for the Maxwell  field from those obtained in previous canonical models.

Finally, in the conclusions we summarize the main results obtained in this
work. Is Lorentz invariance broken when light propagates on a quantum space-time? We
address this question an add our own bias on the subject.

\section{ Phenomenological equations for the propagation of light}

The construction of equations for the propagation of light under the
influence of effects arising from the quantum nature of gravity requires
explicit assumptions about the form that we expect the low energy regime of
these effects will take. In the particular case of cosmological
applications, we would expect that the quantum expectation values should be in consonance with
the properties of the classical metric $g_{ab}$, defining the classical
geometry. For the standard models, this is characterized by a timelike
vector field $t^a$, whose integral lines are the world lines of comoving
observers with 4-velocity $u^a$. We take this as implying that the expectation
values defining the (local) low energy limit should be tensor functions only of
$u^a$, $g_{ab}$ and a scalar function of $t$, where $t$ is an affine parameter
(``time'') for the integral lines of $t^a$.

In view of some recent theoretical results \cite{A-Camelia} we take the
Maxwell tensor $F_{ab}$ as the fundamental physical quantity describing the electromagnetic field. Its components in a local Lorentz
frame are $ F_{0i} = E_i$ and $F_{ij} = \epsilon_{ijk} B_k$, where $E_i$ and
$B_k$ are, respectively, the components of the electric and magnetic field
vectors, and where $\epsilon_{ijk}$ is the Levi-Civita symbol.  We, therefore, do not assume the existence of a vector potential $A_a$.
In the absence of quantum gravity effects, the propagation of light is governed
by the equations for the electromagnetic field $F_{ab}$, which may be written as

\begin{eqnarray}
\label{eq1}
\nabla^a {F^*}_{ab} & = & 0, \\
\label{eq2}
\ \nabla^{a} F_{ab} & = & 4 \pi J_b,
\end{eqnarray}
where ${F^*}_{ab}$ is the dual of $F_{ab}$, $J_a$ is the electric current,
and we recall that the vanishing of the right hand side of (\ref{eq1})
corresponds to the absence of magnetic type currents and monopoles.

The modification of these equations that we envision is the addition of
terms on the right hand sides of (\ref{eq1}) and (\ref{eq2}), corresponding
to the presence of different ``effective currents'' possibly resulting from
quantum gravity effects and to which the electromagnetic field gets coupled. We
shall restrict to terms that are linear in the electromagnetic field, assuming
that non linear effects are of higher order and may be neglected in this
approximation. Our phenomenological equations take the form,

\begin{eqnarray}
\label{eq3}
\nabla^a {F^*}_{ab} & = & \psi_1 t^a F_{ab}+ \psi_2 t^a {F^*}_{ab}
\nonumber
\\
& & +\psi_3 t^a t^c \nabla_c F_{ab}+\psi_4 t^a t^c \nabla_c {F^*}_{ab}
\nonumber \\
& & +\psi_5 t^a t^c t^d \nabla_c [\nabla_d F_{ab}]+\psi_6 t^a t^c t^d
\nabla_c [\nabla_d {F^*}_{ab} ]  \nonumber \\
& & + \psi_7 t^a g^{cd} \nabla_c [\nabla_d F_{ab}]+\psi_8 t^a g^{cd}
\nabla_c [\nabla_d {F^*}_{ab} ], \\
\label{eq4}
\ \nabla^{a} F_{ab} & = & 4 \pi J_b + \chi_1 t^a F_{ab}+ \chi_2 t^a {F^*}_{ab}
\nonumber \\ & & +\chi_3 t^a t^c \nabla_c F_{ab}+\chi_4 t^a t^c \nabla_c
{F^*}_{ab} \nonumber \\
& & +\chi_5 t^a t^c t^d \nabla_c [\nabla_d F_{ab}]+\chi_6 t^a t^c t^d
\nabla_c [\nabla_d {F^*}_{ab} ]  \nonumber \\
& & + \chi_7 t^a g^{cd} \nabla_c [\nabla_d F_{ab}]+\chi_8 t^a g^{cd}
\nabla_c [\nabla_d {F^*}_{ab} ].
\end{eqnarray}
In these equations we are assuming local couplings that lead to expressions
in $F_{ab}$ and its derivatives, and we have included only terms up to second
derivatives. Notice that, from our assumption of a cosmological metric we have,

\begin{eqnarray}  \label{propt}
t^a \nabla_a t^b & = & 0,  \nonumber \\
\nabla^a t^b - \nabla^b t^a & = & 0.
\end{eqnarray}

Similarly, from simple physical arguments, we expect the coefficients
$\chi_i $ and $\psi_i$ to depend only on the scale parameter (``radius of
the Universe'') of the metric, in such a way that they vanish when the
Planck length $\ell_P$ is taken equal to zero. Note that even if some or all
of the coefficients $\chi_i$ and $\psi_i$ are non zero, on observational
grounds they must be small. Thus, we may consider the effect of each term
separately, as any cross effects would be of higher order. This analysis
is carried out in the next subsection.

\subsection{Local Lorentz frame analysis}

The equations are more easily analyzed in a local Lorentz frame, adapted to
the symmetry of the metric. Namely, if the local coordinates are $(x,y,z,t)$, we
have $t^a=(0,0,0,1)$. As a first approximation we also neglect curvature
effects, and equate covariant to ordinary partial derivatives. Moreover, we
assume $J^a=0$, corresponding to ``vacuum'' propagation, and take all $\psi_i$,
and $\chi_i$ as constants.

Using the standard forms for $F_{ab}$ and ${F^*}_{ab}$, we immediately find

\begin{eqnarray}  \label{nablaE}
\nabla \cdot \vec{E} & = & 0, \\
\nabla \cdot \vec{B} & = & 0,
\end{eqnarray}
and
\begin{eqnarray}
\label{geneq1}
\nabla \times \vec{E} + \partial_t \vec{B} & = & \chi_1 \vec{E} + \psi_1
\vec{B} + \chi_2 \partial_t \vec{E} + \psi_2 \partial_t \vec{B} \nonumber
\\
& & + \chi_3\partial_t^2 \vec{E} + \psi_3 \partial_t^2 \vec{B} + \chi_4
\nabla^2\vec{E} + \psi_4 \nabla^2\vec{B}, \\
\label{geneq2}
\nabla \times \vec{B} - \partial_t \vec{E} & = & \chi_5 \vec{B}
+ \psi_5 \vec{E} + \chi_6 \partial_t \vec{B} + \psi_6 \partial_t \vec{E}
\nonumber
\\
& & + \chi_7 \partial_t^2 \vec{B} + \psi_7\partial_t^2 \vec{E}
+ \chi_8 \nabla^2 \vec{B} + \psi_8 \nabla^2 \vec{E}.
\end{eqnarray}

We remark that in (\ref{geneq1}) and (\ref{geneq2}) we have regrouped some
terms from (\ref{eq3}) and (\ref{eq4}), and renamed some constants. Notice that
the factors of $\chi_i$  violate parity conservation. It can also be seen from
(\ref{geneq1}) that the constants $\psi_1, \dots ,\psi_4$, and  $\chi_1, \dots,
\chi_4$ should vanish if one assumes absence of magnetic currents. In this
latter case $F_{ab}$ admits a vector potential $A_a$. This situation will be
analyzed in the next Section.

\subsection{Plane waves}

We look now for plane wave solutions traveling along the z-axis. In this
case, on account of (\ref{nablaE}), we should have,

\begin{eqnarray}  \label{plane1}
\vec{E} & = & \mbox{Re}\left\{(E_x \widehat{e}_x + E_y
\widehat{e}_y)\exp(i(\omega t - k z)) \right\},  \nonumber \\
\vec{B} & = & \mbox{Re}\left\{(B_x \widehat{e}_x + B_y
\widehat{e}_y)\exp(i(\omega t - k z)) \right\}.
\end{eqnarray}

Replacing (\ref{plane1}) in (\ref{geneq1}) and (\ref{geneq2}) we find the
general form for the dispersion relation $k = k(\omega,\chi_i,\psi_i)$. In
view of the smallness of $\chi_i$ and $\psi_i$, this may be expanded up to
linear order in this coefficients, but it will be convenient to keep higher
order terms. If we consider separately terms by their order of derivatives, and
their parity conserving or violating character we find, in the parity conserving
cases:

\begin{eqnarray}
k & = & \sqrt{\omega^2 + 2 i \omega (\psi_1+\psi_5)-4 \psi_1 \psi_5}
\nonumber \\
\label{RD1}
& \simeq & \omega + i(\psi_1+\psi_5), \\
k & = & \omega \sqrt{(1 - 2 (\psi_2 + \psi_6) + 4 \psi_2 \psi_6)}
\nonumber
\\
\label{RD2}
& \simeq & \omega [1 - (\psi_2 + \psi_6) ], \\
k & = & \omega \sqrt{\left[1 -2 i \omega (\psi_3 +\psi_7) - 4 \omega^2
\psi_3\psi_7 \right]}  \nonumber \\
\label{RD3}
& \simeq & \omega [1 - i \omega (\psi_3 +\psi_7) ], \\
k & = & \omega \left[\sqrt{1 - k^2 (\psi_4 -\psi_8)^2} + i k
(\psi_4+\psi_8)\right]^{-1} \nonumber \\
\label{RD4}
& \simeq & \omega [1 - i \omega (\psi_4 +\psi_8) ],
\end{eqnarray}
and for the parity violating cases:
\begin{eqnarray}
\label{RD5}
k & = & \sqrt{\omega^2 \pm 2 k (\chi_1 -\chi_5) +4 \chi_1 \chi_5)}
\nonumber
\\
& \simeq & \omega \pm (\chi_1 -\chi_5), \\
\label{RD6}
k & = & \omega \left[\sqrt{1 - (\chi_2+\chi_6)^2} \pm i ( \chi_2- \chi_6)
\right]  \nonumber \\
& \simeq & \omega [1 \pm i ( \chi_2- \chi_6)], \\
\label{RD7}
k & = & \omega \left[\sqrt{1 + \omega^2 (\chi_3+\chi_7)^2} \pm \omega(
\chi_3- \chi_7) \right]  \nonumber \\
& \simeq & \omega [1 \pm \omega (\chi_3- \chi_7)], \\
\label{RD8}
k & = & \omega \left[1 \pm 2 k (\chi_4 -\chi_8) - 4 k^2 \chi_4
\chi_8\right]^{-1/2}  \nonumber \\
& \simeq & \omega [ 1 \pm \omega (\chi_4 -\chi_8)].
\end{eqnarray}

If we use these results to obtain the corresponding amplitudes, we find
that in the parity conserving cases there is a single mode of propagation,
and the polarization of a plane polarized wave is preserved in time, while
in the parity violating cases there are two modes, one corresponding to
right, and the other the left circular polarization. Possible
observational effects for these cases have been discussed in \cite{GleiKo}.

Considering now particular cases, we see that (\ref{RD1}) corresponds to a
frequency independent amplification or attenuation (depending on the sign
of $\psi_1+\psi_5$) of the wave amplitude with time, with no effects on the
polarization. The effect would be absent for $\psi_1= -\psi_5$, but this
would still leave a term quadratic in $\psi_1$, that would behave as a
tachyonic mass added to the photon. The important question here is the
order of magnitude of these couplings. In the absence of a theory we may only
conjecture that we would expect these to be of the order of the corresponding
dimensional quantities that characterize the model. In our case these are
the Planck length $\ell_P$, and time $t_P$ and possibly, for instance, the
radius (or scale) $a(T)$, and age $T$ (or the Hubble constant $H = \dot{a}
/a $) of the Universe. Then, since $\psi_1$ and $\psi_5$ should have
dimension [length]$^{-1}$, one would
expect these quantities to be  of the order of $t_P H^2$, which, after
multiplication by the time to travel through cosmological distances, is too
small to have any phenomenological relevance.

In the case (\ref{RD2}) we have a frequency independent change in the
velocity of the waves, which would be larger or smaller than that of light,
depending on the sign of $\psi_2 + \psi_6$. Since $\psi_2$ and $\psi_6$ are
dimensionless, an estimate for them could be $t_P H$, which is again too
small for observable consequences.

The case (\ref{RD4}) is conceptually similar to (\ref{RD1}), leading again to
amplification or attenuation of the waves. However, in this case $\psi_4$ and
$\psi_8$ have dimension of $[length]$. Then, if we take them to be of the order
of $\ell_P$, the relevant quantity would be of order $\ell_P
L/\lambda^2$, where $\lambda$ is wavelength of the wave and $L$ of the order of
the distance to cosmological sources. Perhaps surprisingly, if we assume
$\lambda = 10^{-5} cm$, (for visible light), and $L=10^9 ly$, we find already
$\ell_P L/\lambda^2 \simeq 10^{3}$. This would imply that visible light would not reach us but, in fact, not only visible light but also
$\gamma$-rays are observed from cosmological sources. This implies that
either $\psi_4$ and $\psi_8$ are much smaller than this scale or $\psi_4
\simeq -\psi_8$. If we take $\psi_4 = -\psi_8$, we find $k \simeq \omega -2
\psi_8^2 \omega^3$. This implies a group velocity $v_g = 1- 6 \psi_8^2
\omega^2$ which, even for high energy $\gamma$-rays and cosmological times,
gives an effect that is too small for observable consequences.

The case (\ref{RD3}) is similar to (\ref{RD4}), but here we might question
if it is acceptable to have second order time derivatives on the right
hand side of the equation, or we should disregard this possibility.

Similar considerations regarding orders of magnitude and observability
hold for the parity violating cases (\ref{RD5}), (\ref{RD6}), (\ref{RD7}), and
(\ref{RD8}). The last case, with $\chi_4=- \chi_8$, was obtained in \cite{GaPu}.
We notice however that the available observational data indicates that  $\chi_4 \simeq 10^{-4}$\cite{GleiKo}, much smaller than the expected value of order one.

It is worth mentioning that the same observational data can be used to test the
validity of a recent result that also predicts a rotation of the
polarization vector \cite{Thiemann}. Denoting by $D$ the cosmological distance to the source of
the incoming radiation, by $\theta_{GP}=\chi \ell_P k^2 D$ and $\theta_{ST}=
\ell_P^{2\alpha} L^{1- 2\alpha}k^2 D$ the rotation angle of the polarization direction
obtained by Gambini-Pullin and Sahlmann-Thiemann respectively, and comparing the
uncertainty of these angles with the observational data, one obtains the
following restriction for the constants $\alpha$ and $L$,

$$
\left(\frac{\ell_P}{L}\right)^{2\alpha-1}\simeq \chi \simeq 10^{-4}.
$$

However, these constants must satisfy the following inequalities $0<
\alpha<\frac{1}{2}$ and $L>>\ell_P$\cite{Thiemann}, which clearly violate the
above equation. This would suggest that the kinematical states used by the authors to
derive their equations cannot be considered as semiclassical states.

\section{The interaction between the gravitational and electromagnetic fields}

In this section we first review the standard classical Lagrangian and Hamiltonian formulation of coupled Maxwell and gravitational fields. This review is then used to argue that, even when we promote the fields to quantum operators, there will always be relations that remain valid for the quantum theory.

 The maxwell field on a curved spacetime is given by an exact 2-form 
$$F= dA,$$
 with $A$ the Maxwell connection 1-form. Given a local coordinate system with timelike and spacelike coordinate vectors $e_o^a$ and $e_i^a$ respectively, we define the electric and magnetic fields as $E_i = F_{io}$ and $B_{i} =\epsilon _{ijk}F_{jk}$, where $\epsilon_{ijk}$ is the Levi-Civita symbol. With these definitions (and assuming that $A_o=0$ since we are only concerned with propagating waves) the relations between the fields and the potential can be written as 
\begin{equation}
E_i=-\partial_t A_i,\label{E}
\end{equation}
 and 
\begin{equation}
B_i=\epsilon _{ijk}\partial_j A_k. 
\end{equation}
Note that these relations do not depend on any metric and are just the coordinate components of the above equation.

Assume now that we give any Lagrangian density $\mathcal{L}=\mathcal{L}(F,g)$ describing the coupling between the Maxwell and gravitational field. By definition, the Maxwell field will satisfy
$$dF=0,$$
since it is not part of the Euler-Lagrange equations of motion. The coupling between the metric and electromagnetic fields will of course depend on the detailed form of the Lagrangian density. In particular, for 
\begin{equation}
\mathcal{L_o}=-\frac{1}{2}F\wedge {}^*F \label{L}
\end{equation} 
we obtain the standard field equations
$$d ^* \! F=0,$$
but we will allow for more general types of couplings that might arise from the semiclassical aproximation when we promote the classical fields to quantum operators.
To construct a Hamiltonian formulation we first define the canonical momentum conjugated to $A_i$
$$\pi^i = \frac{\partial \mathcal{L}}{\partial(\partial_t A_i)}$$
and then invert this relation to obtain
\begin{equation}
\partial_t A_i = G_i(\pi^j,A_k).\label{G}
\end{equation}
The Legendre transformation $\mathcal{H}=\pi^iG_i - \mathcal{L}$ then gives the desired Hamiltonian density. Note that, by construction, the hamilton equation of motion for $A_i$ will be (\ref{G}). Note also that, regardless of the form of $G_i$, $E_i$ must satisfy equation (\ref{E}). Thus, we expect that for a non-standard interaction Hamiltonian, $-E_i$ will not be the conjugate momentum to $A_i$.

If we now promote the classical fields to quantum operators $(\hat{A},\hat{F},\hat{g})$ we might expect several changes in the quantum hamiltonian but the basic definitions should remain valid. For example, 
$$\hat{F}= d\hat{A},$$
says that $\hat{F}$ is still the curvature of $\hat{A}$. Thus, the equation 
$$d\hat{F}=0,$$
should hold as an identity for $\hat{A}$. Moreover, the above relation does not depend on $\hat{g}$, therefore even if we must regularize the metric operator it should remain unchanged. Even if we take expectation values of this relation and assume the state is a direct product of a coherent state for the Maxwell field and a ``semiclassical state" for the gravitational field one should reobtain 
$$F_{class}= dA_{class}, \hspace{1cm}dF_{class}=0,$$
at least at a semiclassical approximation. Thus, the plan is as follows:
\begin{enumerate}
\item We start by impossing $\hat{F}= d\hat{A}$.
\item We then take the regularized energy density of the electromagnetic field written in terms of $E_i$ and $B_i$ and  write a modified relation between $E_i$ and $\pi^i$ so that, via the hamilton equations,  the relation 
$$E_i= - \partial_t A_i$$
holds order by order in a perturbation expansion.
\item With the desired relation $E_i = -G_i$ we obtain the second equation of motion for $\pi^i$ and then rewrite the final result as a modified set of equations for the Maxwell fields.
\end{enumerate}

\subsection{The phenomenological Hamiltonian and Lagrangian densities}

Following the Gambini-Pullin approach\cite{GaPu}, the phenomenological interaction Hamiltonian density (obtained by taking expectation values of the regularized quantum hamiltonian with coherent states for the Maxwell field and ``weave states" for the gravitational field) is given by,

\begin{equation}
\mathcal{H_{EB}}=\frac{1}{2}\left(\vec{E}^2+\vec{B}^2\right)+\chi
l_p\left(\vec{E}\cdot\vec{\nabla}\times\vec{E}+\vec{B}
\cdot\vec{\nabla}\times\vec{B}\right).\label{hamiltoniano}
\end{equation}

where $\chi$ is a phenomenological coupling constant that destroys parity invariance.

The relationship between the fields and the conjugate variables $(\vec{A},\vec{\pi})$ is, by assumption, given by
\begin{eqnarray}
\vec{B}&=&\vec{\nabla}\times\vec{A}\\
\vec{E}&=&-\vec{\pi}+ \chi l_p \vec{F}[\vec{\pi}],
\end{eqnarray}
where we have adopted the vectorial notation for ease of writing and where $\vec{F}[\vec{\pi}]$ is a function of $\vec{\pi}$ and its derivatives. If we now insert these relations in the above Hamiltonian and impose that $\vec{E} = -\partial_t \vec{A}$ (which must be true by definition), the functional $\vec{F}$ is determined in a unique way via the Hamilton equation for $\vec{A}$, leading to
\begin{equation}
\vec{E} =  -\vec{\pi}+ 2 \chi l_p \vec{\nabla}\times\vec{\pi}.\label{E-P}
\end{equation}
This allows us to obtain the second equation of motion for $\vec{\pi}$. Hamilton equations for the conjugate variables then read
\begin{eqnarray}
\partial_t\vec{A} & = & \vec{\pi}- 2 \chi l_p \vec{\nabla}\times\vec{\pi},\\
\partial_t\vec{\pi} & = & -\vec{\nabla}\times\vec{\nabla}\times(\vec{A}+2\chi l_p\vec{\nabla}\times\vec{A}).
\end{eqnarray}

Combining these equations and assuming the Coulomb gauge we obtain 

\begin{equation}
(\partial_t^2 - \nabla^2)\vec{A} = \mathcal{O}\left((\chi l_p)^2\right).
\end{equation}
Alternatively, we can derive Maxwell's equations for the electric and magnetic fields yielding 
\begin{eqnarray}
 \label{Maxwell} \nabla\times\vec{B} - \partial_t\vec{E} & = & \mathcal{O}\left((\chi l_p)^2\right),\\ 
\label{Maxwellp} \nabla\times\vec{E} + \partial_t\vec{B} & = & \mathcal{O}\left((\chi l_p)^2\right),\\   
 \nabla\cdot\vec{E} & = & \mathcal{O}\left((\chi l_p)^2\right),\\
\nabla\cdot\vec{B} & = & 0.\label{divB}
\end{eqnarray}

{\bf Remarks:}

\begin{itemize}
\item Note that our field equations are Lorentz invariant and have a standard dispersion relation.
\item Starting from the same eq. (\ref{hamiltoniano}), Gambini and Pullin\cite{GaPu} assumed a different relation between the electric field and the momentum, namely $\vec{E}=-\vec{\pi}$. This assumption leads to the following equations of motion for the conjugate pair, 
\begin{eqnarray}
\partial_t\vec{A} & = & -\left(\vec{E}+2\chi
 l_p\nabla\times\vec{E}\right),\label{H1}\\
\partial_t\vec{E} & = & \nabla\times\left(\vec{B}+2\chi l_p
 \nabla\times\vec{B}\right).\label{H2}
\end{eqnarray}
In terms of the fields, the equations can be rewritten as
\begin{eqnarray}
\nabla\times\vec{B} - \partial_t\vec{E} & = & 2 \chi l_p \nabla^2\vec{B},\label{G-P1}
\\ \nabla\times\vec{E} + \partial_t\vec{B} & = & -2 \chi l_p \nabla^2\vec{E}, \label{G-P2}
\end{eqnarray}
which, unlike the ones we obtained, are not Lorentz invariant. It is important to notice that this breakdown of invariance follows from the assumption that $-\vec{E}$ and $\vec{A}$ are canonical variables. Although this is the case for the classical theory, it might not remain valid when we promote the fields to quantum operators. On the other hand, the quantum version of $\vec{E} = -\partial_t \vec{A}$ should remain valid since this is just a consequence of $F = dA$. We therefore conjecture that the homogeneous Maxwell equations, contained in $F = dA$, should remain unchanged for the full theory. 
\item It is worth mentioning that using a completely different quantization procedure Ellis et. al. obtain a generalized set of Maxwell equations that keep the homogeneous part unchanged\cite{Ellis}. 
\end{itemize}

It is also possible to derive, up to linear order in $\chi$, the Lagrangian density associated to the Hamiltonian (\ref{hamiltoniano}), via the Legendre transformation $\mathcal{L}=\pi^i\partial_tA_i - \mathcal{H}$. If we invert (\ref{E-P}) we get
\begin{equation}
\vec{\pi} =  -\vec{E}- 2 \chi l_p \vec{\nabla}\times\vec{E},
\end{equation}
which leads to the following Lagrangian density
\begin{equation}
\mathcal{L_{EB}}=\frac{1}{2}\left(\vec{E}^2-\vec{B}^2\right)+\chi
l_p\left(\vec{E}\cdot\vec{\nabla}\times\vec{E}-\vec{B}
\cdot\vec{\nabla}\times\vec{B}\right).
\end{equation}
The Euler-Lagrange equation of motion for the potential can be written in terms of the fields as: 
\begin{equation}
\nabla\times\vec{B} - \partial_t\vec{E}  = - 2 \chi l_p\nabla\times\left(\nabla\times\vec{B} - \partial_t\vec{E}\right) + \mathcal{O}\left((\chi l_p)^2\right),\\ \label{Maxwell2} 
\end{equation}
Note also that in principle the above equations could give more solutions
than plane waves since they are second order PDE's. Thus, there seems to be more solutions to the above equations than Hamilton´s equations (\ref{Maxwell}), since the ones obtained directly from the Hamiltonian do not contain terms linear in $\chi$. The extra terms appearing in (\ref{Maxwell2}) arise when we invert the relation between $\vec{E}$ and $\vec{\pi}$.
We can see that both equations contain the same solutions, in the linear approximation, by the following consideration. Let $\vec{C}=\vec{\nabla}\times\vec{B} - \partial_t\vec{E}$. The solution to Hamilton equations correspond to $\vec{C}=0$. We now search for non trivial solutions to
\begin{equation}
 2 \chi l_p\vec{\nabla}\times\vec{C} + \vec{C}=0,\\ \label{C} 
\end{equation}

Note that if $\vec{C}$ satisfies
(\ref{C}), then it is also a solution of
\begin{equation}
\label{eqE2}
\vec{C}+(2\chi l_p)^2 \nabla^2 \vec{C} =0,
\end{equation}
and this implies that solutions of (\ref{C}) have a ``monochromatic''
(spatial) Fourier spectrum with $k =(2\chi l_p)^{-1}$. Such contributions with
wavelength of the order of the Planck length must be absent if the low energy
approximation (characteristic lengths much larger than $l_p$ ) is to be consistent.

Assuming this cut off is implemented, the only solution consistent with our approximation is $\vec{\nabla}\times\vec{B} - \partial_t\vec{E}=0$.

Using a somewhat different approach, L. Urrutia\cite{Urrutia} developed a Lagrangian formulation that leads to equations (\ref{G-P1}, \ref{G-P2}). Starting from (\ref{L}) and defining $\vec{E}$ by Eq. (\ref{H1}) and $\vec{B}$ as the curl of $\vec{A}$, Urrutia obtains a Lagrangian density for $\vec{A}$ such that the corresponding Euler-Lagrange equations
yield (\ref{G-P1}, \ref{G-P2}). This Lagrangian, however, is a non-local functional of $\vec{A}$. The fact that a Lorentz violating dispersion relation
$\omega_{\pm}=\sqrt{k^2\mp 4\chi l_pk^3}\simeq  \vert k\vert (1\mp2\chi
l_p\vert k\vert)$ is found for the set of equations (\ref{G-P1}, \ref{G-P2}) is
consistent with a recent result advocating that non-local Lagrangians can
generate non-Lorentz-invariant dispersion laws\cite{Lehnert}.
The non-local Lagrangian arose, however, from an assumed non-local
relation between $\vec{E}$ and $\vec{A}$ which will not be true if one follows our assumption. 

Unlike the approach of Urrutia\cite{Urrutia} the Lagrangian density that yields eqs.(\ref{Maxwell},\ref{Maxwellp}) is a local functional of the potential.
This result is in agreement with recent work of J. Bros and H.
Epstein \cite{B-E} which states that microcausality and energy positivity in all
frames imply Lorentz invariance of dispersion laws. (Within the domain of our
assumption the energy spectrum will be positive definite since the Hamiltonian
can be written as $\mathcal{H_{EB}}=\frac{1}{2}\left(\vec{E}+\chi
l_p \vec{\nabla}\times\vec{E}\right)^2+ \frac{1}{2}\left(\vec{B}+\chi
l_p \vec{\nabla}\times\vec{B}\right)^2$ plus terms of order $\chi^2$).

It would appear that the issue of whether or not Lorentz invariance is
broken depends on the specific relation between the fields and the potential.
The local Lagrangian constructed from the potential and its derivatives
yields Lorentz invariant dispersion laws, whereas the non-local Lagrangian
breaks the invariance. Following our assumption that $\hat{F} = d\hat{A}$ we conjecture that the quantum fields will be local functionals of $\hat{A}$.

In principle, if the Hamiltonian were given as a perturbation expansion in $\chi$, it would be possible to introduce higher order corrections to the relationship between the electric field and conjugate momentum, eq. (\ref{E-P}), so that Faraday's law is preserved at each order in that expansion.

\section{Final comments and conclusions}

We first summarize the main results of this work. In Section 2 
we concentrate on the phenomenological field equations for Maxwell fields that could
arise from the interaction with the gravitational field in the
semiclassical approximation. We show that, up to linear order, there is just one acceptable way to obtain a non-standard dispersion relation, namely, field equations with parity
violating coupling constants\cite{GaPu}.

Is the universe living in a state of definite parity? Voting for the
affirmative, the cosmological nature and detailed temporal structure of gamma
ray bursts could serve as a tool to observe the predicted deviations from the
standard dispersion relations. As was shown on a previous work\cite{GleiKo} the
phenomenological constant is rather small and future observations could be
useful to decide if indeed it is non-vanishing. However, using the available data we can essentially rule out the Sahlmann-Thiemann construction.

In view of the present observational data for cosmological sources and the
ambiguities of the available models for the interaction of Maxwell fields
with a semiclassical space-time it is fair to ask ourselves whether or not
we should still have faith in Lorentz invariance. In Section 3 we show that a very natural assumption leads to Lorentz invariant field equations. Our conclusion is that
we have no reason to believe that a quantum theory of gravity
would change the invariance and thus conjecture that Lorentz invariance will still hold in the quantum interaction between gravitons and photons.

\ack

This work was supported in part by grants of the National University of
C\'ordoba and Agencia C\'ordoba Ciencias. (Argentina). R.J.G. and C.N.K.
are members of CONICET. F.P. is supported by SECYT - UNC.


\end{document}